\documentclass[twocolumn]{aastex631}

\usepackage{amsmath}

\begin{document}

\title{TRIANGULUM IV: A POSSIBLE ULTRA-DIFFUSE SATELLITE OF M33}

\correspondingauthor{Itsuki Ogami}
\email{itsuki.ogami@grad.nao.ac.jp}

\author[0000-0002-0786-7307]{Itsuki Ogami}
\affiliation{The Graduate University for Advanced Studies (SOKENDAI), 2-21-1 Osawa, Mitaka, Tokyo 181-8588, Japan}
\affiliation{National Astronomical Observatory of Japan, 2-21-1 Osawa, Mitaka, Tokyo 181-8588, Japan}

\author[0000-0002-3852-6329]{Yutaka Komiyama}
\affiliation{Department of Advanced Sciences, Faculty of Science and Engineering, Hosei University, 3-7-2 Kajino-cho, Koganei, Tokyo 184-8584, Japan}

\author[0000-0002-9053-860X]{Masashi Chiba}
\affiliation{Astronomical Institute, Tohoku University, Aoba-ku, Sendai, Miyagi 980-8578, Japan}

\author{Mikito Tanaka}
\affiliation{Department of Advanced Sciences, Faculty of Science and Engineering, Hosei University, 3-7-2 Kajino-cho, Koganei, Tokyo 184-8584, Japan}

\author[0000-0001-8867-4234]{Puragra Guhathakurta}
\affiliation{Department of Astronomy and Astrophysics, University of California Santa Cruz, University of California Observatories, 1156 High Street, Santa Cruz, CA 95064, USA}

\author[0000-0001-6196-5162]{Evan N. Kirby}
\affiliation{Department of Physics and Astronomy, University of Notre Dame, Notre Dame, IN 46556, USA}

\author[0000-0002-4013-1799]{Rosemary F.G. Wyse}
\affiliation{Department of Physics and Astronomy, Johns Hopkins University, Baltimore, MD 21218, USA}

\author[0000-0001-5522-5029]{Carrie Filion}
\affiliation{Department of Physics and Astronomy, Johns Hopkins University, Baltimore, MD 21218, USA}

\author[0000-0001-6503-8315]{Takanobu Kirihara}
\affiliation{Kitami Institute of Technology, 165, Koen-cho, Kitami, Hokkaido 090-8507, Japan}

\author[0000-0003-4656-0241]{Miho N. Ishigaki}
\affiliation{National Astronomical Observatory of Japan, 2-21-1 Osawa, Mitaka, Tokyo 181-8588, Japan}

\author[0000-0002-8758-8139]{Kohei Hayashi}
\affiliation{National Institute of Technology, Sendai College, Natori, Miyagi 981-1239, Japan}
\affiliation{Astronomical Institute, Tohoku University, Aoba-ku, Sendai, Miyagi 980-8578, Japan}
\affiliation{Institute for Cosmic Ray Research, The University of Tokyo, Kashiwa, Chiba 277-8582, Japan}



\begin{abstract}

We report the detection of a dwarf satellite candidate (Triangulum IV: Tri IV) of the Triangulum galaxy (M33) using the deep imaging of Subaru/Hyper Suprime-Cam (HSC). From the apparent magnitude of the horizontal branch in Tri IV, the heliocentric distance of Tri IV is estimated to be $932_{-43}^{+49}$ kpc, indicating that Tri IV is located at the distance of $75_{-40}^{+48}$ kpc from the M33 center. This means that Tri IV is the probable satellite of M33, because its distance from M33 is within the virial radius of M33. We also estimate its surface brightness of $\mu_{\it V} = 29.72_{-0.10}^{+0.10}$ ${\rm mag~arcsec}^{-2}$, and half-light radius of $r_h = 1749_{-425}^{+523}$ pc, suggesting that Tri IV is an ultra-diffuse galaxy or dynamically heated galaxy. The surface brightness of Tri IV is too low to be detected in the previous survey, so this detection suggests that much fainter satellites may be present in the outskirts of M33.

\end{abstract}


\section{Introduction} \label{section:intro}

In the Lambda-dominated cold dark matter ($\Lambda$CDM) scenario, large spiral galaxies have been formed by the accretion and merging of the small stellar systems \citep[e.g.,][]{white1991}. The accreted systems are tidally disrupted, and their stellar components accumulate in the outskirts of the spiral galaxies. These accumulated structures consist of smooth structures (stellar halo) and stellar substructures (stellar streams and/or surviving satellite galaxies). This scenario successfully explains the large-scale structure in the universe ($>$ 1 Mpc), but several discrepancies with observational results reveal on a small scale ($<$ 1 Mpc). One of the problems is the `missing satellites problem' \citep[e.g.,][]{moore1999}. Cosmological {\it N}-body simulations have predicted that $\Lambda$CDM halos of the Milky Way (MW) and M31 size galaxies host 100–1000 subhalos, but only dozens of satellite galaxies have been discovered in these galaxies. With the recent discovery of ultra-faint galaxies (UFDs) by large-scale telescopes \citep[e.g.,][]{willman2005,belokurov2006a,homma2024}, this problem is gradually being resolved in massive galaxies. However, to date UFDs have only been detected in a few, nearby large spiral galaxies, so it is not known whether this problem remains in other large galaxies and/or other mass-scale galaxies.

The missing satellites problem arises not only in massive spirals but also in intermediate-mass spiral galaxies such as the Triangulum galaxy (M33). M33 is the third most massive spiral galaxy in the Local Group, with a stellar mass of $3\times10^9~M_{\odot}$ \citep{mcconnachie2012} and a halo mass of $\sim 10^{11}~M_{\odot}$ \citep{corbelli2014}. With this mass scale, simulation studies have predicted that M33-like galaxies possess satellite galaxies: in cosmological simulations, such galaxies are expected to have a dozen of satellite galaxies with luminosity $> 10^4~{\rm L}_\odot$ \citep{dooley2017,patel2018}. However, only two satellite galaxies have been identified in M33 so far: Andromeda (And) XXII/Triangulum (Tri) I \citep{martin2009,chapman2013}, and Pisces (Pis) VII/Tri III \citep{collins2024}.

In recent years, analyses of deep imaging of the outskirts of M33 have suggested that M33 has a stellar halo, likely formed through the accretion of satellite galaxies. Hubble Space Telescope (HST) multicolor observations found the existence of the stellar halo with a power-law index $\alpha \sim -3$ in the central region \citep[$<$ 5 kpc;][]{smercina2023}. In addition, the observations using Hyper Suprime-Cam (HSC) mounted on the Subaru Telescope identified the stellar component in the outer region ($>$ 10 kpc) with $\alpha < -3$, which is possibly the outer part of the stellar halo \citep{ogami2024}. These results indicate the existence of the stellar halo in M33, which is formed by the accretion of satellite galaxies, thereby suggesting the presence of hidden satellite galaxies in the outskirts of M33. 

In order to unravel the missing satellites problem in M33, we conduct a wide and deep survey of M33 using Subaru/HSC \citep{miyazaki2012,furusawa2018,komiyama2018a}. HSC is an excellent instrument for high-quality detection and characterization for the diffuse structure using the faint red giant (RGB), and horizontal branch (HB) stars as tracers. HSC’s three broadband filters, {\it g}-/{\it r}-/{\it i}-bands, and one narrowband filter, {\it NB515}, are the best combination to remove contaminants. By combining the three broadband filters, it is possible to remove unresolved background galaxies using the color of objects \citep[e.g.,][]{fukushima2019,ogami2024}. {\it NB515}, centered on the MgH/Mgb lines which are sensitive to stellar surface gravity \citep[e.g.,][]{majewski2000}, can be used to remove foreground galactic dwarf stars \citep{komiyama2018,ogami2024a}. Therefore, the removal of these contaminants and the wide and deep HSC observations are expected to unveil the nature of the faint M33 stellar halo and to detect its faint substructures.

In this letter, we report the discovery of a dwarf satellite galaxy, Triangulum IV (hereafter, Tri IV), in the western outskirts of M33. This letter is organized as follows. In Section \ref{section:observation}, we present our imaging with Subaru/HSC and the method for data analysis. We introduce our detection of Tri IV in Section \ref{section:detection} and derive its photometric properties in Section \ref{section:properties}. Finally, we discuss the implications of our findings and present our conclusions in Section \ref{section:conclusions}.

\section{Observation and Data Reduction} \label{section:observation}

We observed the western region of M33 in the {\it g}-, {\it r}-, and {\it i}-bands and {\it NB515} using Subaru/HSC during the nights of 2022 and 2023 (PI: I. Ogami; Proposal ID: S22B-107, S23B-072) with the seeing ranged from 0\farcs50 to 1\farcs1. These observations are part of a survey to map the M33 outer region with seven HSC pointings. In this letter, we focus on the western region of our survey which is the same region as in \citet{ogami2024}. 

The observed raw images are processed and calibrated using the HSC pipeline \citep[version 8.4; hscPipe;][]{bosch2018}, which is based on a software suite being developed for the Vera C. Rubin Observatory data \citep{axelrod2010,juric2017,ivezic2019} project. We perform the reduction, the calibration, the extinction correction, and the artificial star tests, using the same method as described \citet{ogami2024a}. After these procedures, the selection of the stellar sources is performed in the same process as in \citet{ogami2024}. Briefly, `pure' point sources are selected by using the `extendedness' parameter provided by hscPipe and performing color selection in the $r-i$ v.s. $g-r$ diagram. Full details of the observations and data reduction are presented in \citet{ogami2024}.

\begin{figure*}[ht!]
\includegraphics[width=2\columnwidth]
{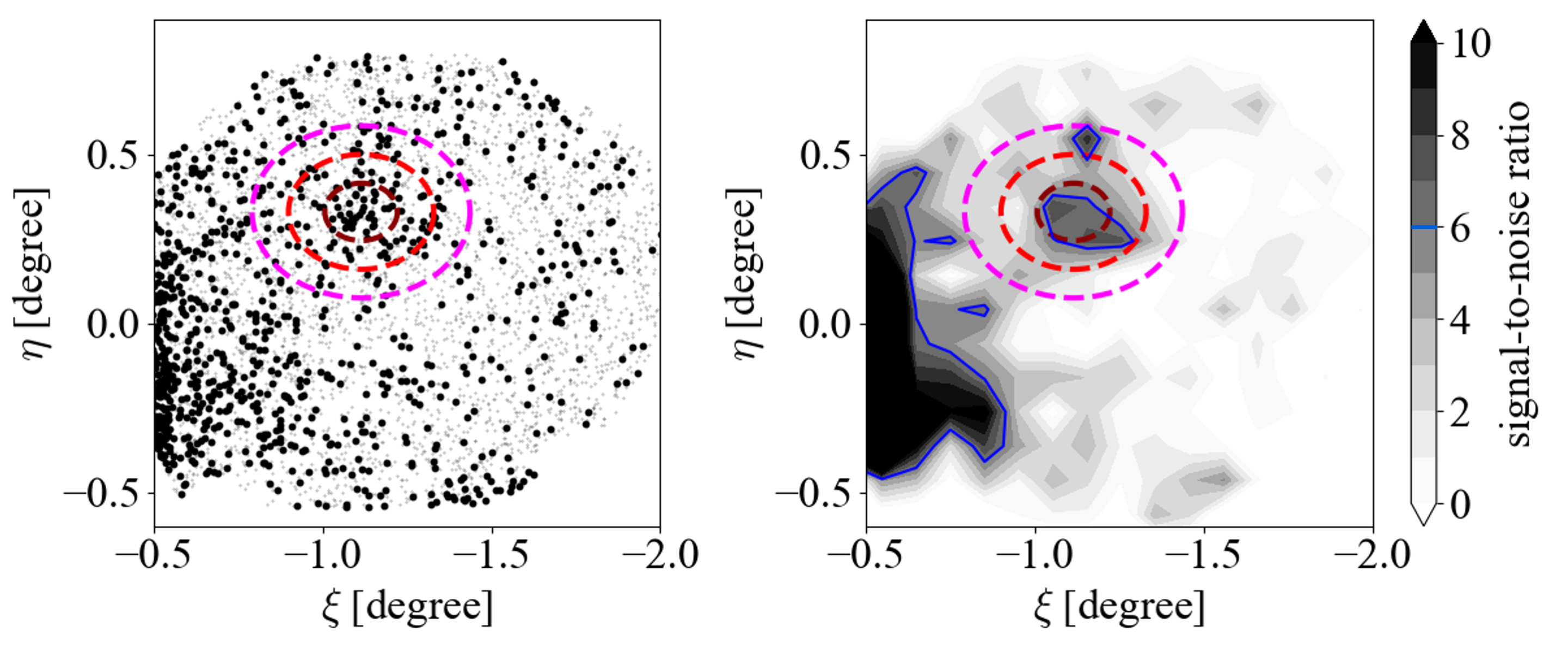}
\caption{Left: The spatial distribution of the `pure' point sources with $p_{\rm RGB}>0.9$. The dark-red, red and pink ellipse show the region within 1, 2, and 3 times the half-light radius based on the estimated parameter (see Section\ref{section:properties}). Right: the signal-to-noise (S/N) ratio map. We assume that the region 0.3 degree below the center of the detected overdensity is the control M33 field, and we calculate the signal-to-noise ratio using the method in \citet{yang2023}. The blue line represents a contour with S/N=6.
\label{fig:SpatialDistribution}}
\end{figure*}

\section{Detection of Triangulum IV}\label{section:detection}

Since M33 is located at low galactic latitude, foreground disk/halo stars are the main contaminants for the brightness range of RGB stars in M33. To eliminate the foreground contamination sources, we use {\it g-}/{\it i-}bands and {\it NB515}. The elimination method is the same as \citet{ogami2024a}, and we conduct this method to extract RGB stars in the M33 halo. Briefly, using `pure' point sources extracted from the position on the {\it r-i} v.s.  {\it g-r} diagram, RGB stars are selected by deriving the probability $p_{\rm RGB}$ of a star being an RGB star from its position on the {\it NB515-g} v.s. {\it g-i} diagram with color and magnitude cut.

Using the selected catalogs, we construct the spatial distribution of RGB stars with $i_0<23.5$ weighted by $p_{\rm RGB}$, and we can see a few high-significance peaks. According to \citet{ogami2024a}, it is difficult to estimate the contamination of background galaxies prescisely, so we first perform an exploration using only RGB stars. A number of spatial overdensities of RGB stars are observable (see Figure \ref{fig:SpatialDistribution}), with some peaks corresponding to noise artifacts. We visually inspect the apparent color-magnitude distributions of each overdensity to identify and remove such spurious signals. After the visual inspection, we confirm one overdensity of RGB stars at $(\xi,\eta) \sim (-1.1,0.4)$, where $(\xi,\eta)$ is the standard coordinate centered on M33. The left panel of Figure \ref{fig:SpatialDistribution} shows the spatial distribution of `pure' point sources with $p_{\rm RGB} > 0.9$, and the right panel is the map of the signal-to-noise ratio (S/N). We assume that the region 0.3 deg below the center of this overdensity is the control M33 field (i.e., smooth halo without substructures), and we calculate S/N based on the equation of \citet{yang2023}. This map indicates that S/N of this overdensity exceeds 6. This means that this region has six times as many `pure' point sources as the adjacent region. Besides this, the overdensities from noise artifacts have lower S/N in the right panel of Figure \ref{fig:SpatialDistribution}. There are no corresponding celestial objects in the NASA/IPAC Extragalactic Database or Simbad, so we designate it as Triangulum IV (Tri IV). In Section \ref{section:properties}, the structural parameters such as the half-light radius, $r_h$, of Tri IV are derived, using the black dots in Figure \ref{fig:SpatialDistribution}. In Figure \ref{fig:SpatialDistribution}, the dark-red, red, and pink ellipses indicate the 1, 2, and 3 times $r_h$.

\begin{figure*}[ht!]
\includegraphics[width=2\columnwidth]
{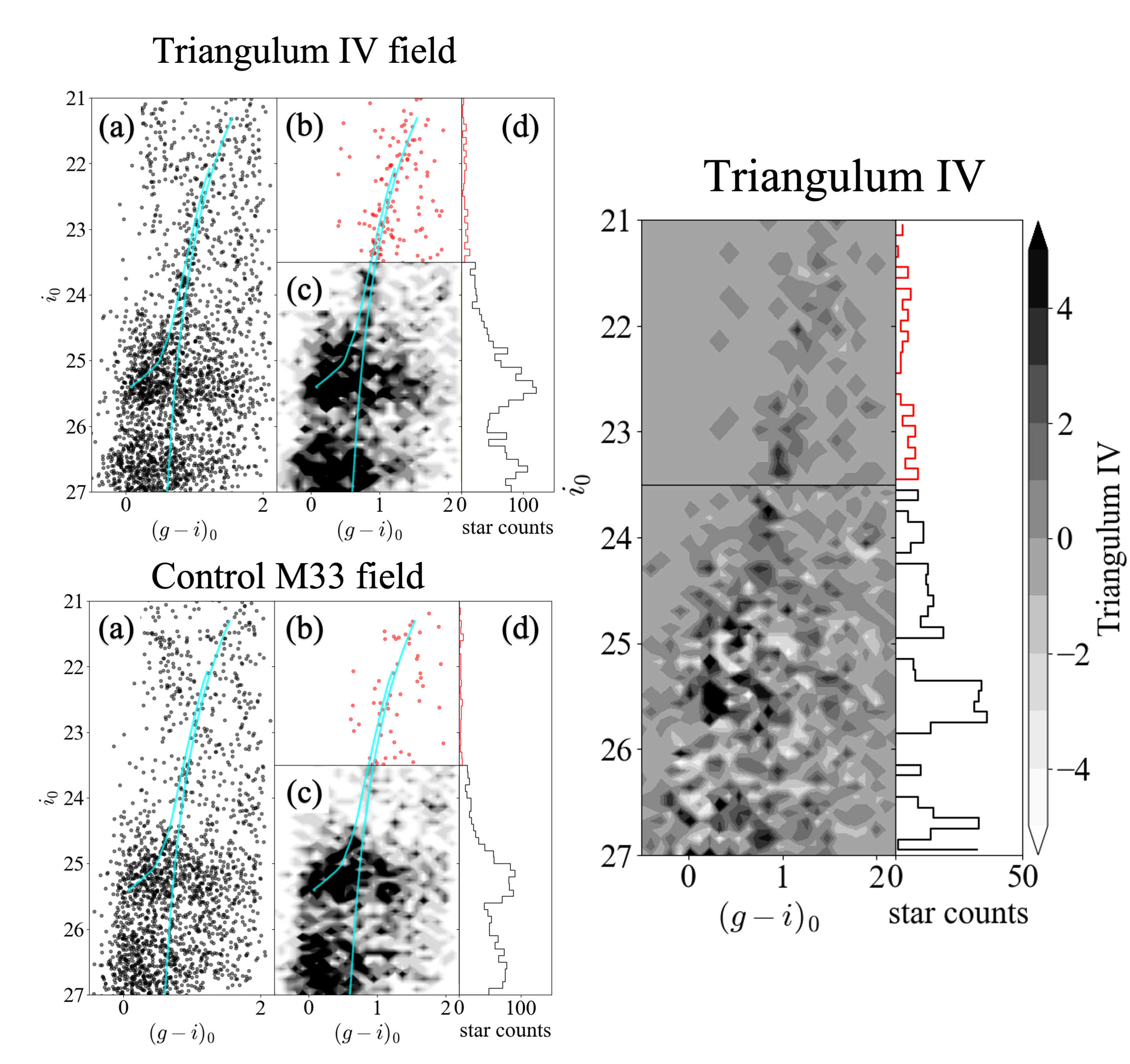}
\caption{Upper left: The color-magnitude diagram and luminosity function within 3 times the half-light radius of Tri IV. All `pure' point sources are used in the panel (a), and the stars with $p_{\rm RGB}>0.9$ are used in the panel (b). The panel (c) is a Hess diagram of the panel (a). The panel (d) shows the luminosity function of the stars with $p_{\rm RGB}>0.9$ (red; $i_0<23.5$) and all `pure' point sources (black; $i_0>23.5$). The cyan lines indicate the BaSTI isochrone \citep{hidalgo2018,pietrinferni2021} with 10 Gyr and $[\mathrm{Fe/H}] = -1.90$ assumed the heliocentric distance of 958 kpc. Lower left: The same as in the upper left figure but for the diagrams for the control M33 field. Right: Control M33 field subtracted Hess diagrams and luminosity function. The color bar of the Hess diagram indicates the difference between the Tri IV field (upper left figure) and the control M33 field (lower left figure), indicating the true intensity of Tri IV.
\label{fig:CMD}}
\end{figure*}

Figure \ref{fig:CMD} shows the color-magnitude diagram (CMD) of stars within 3 times $r_h$ of Tri IV (see Section \ref{section:properties}) and the CMD of the control M33 field with the same area but offset by 0.3 degree toward the south. The panel (a) in Figure \ref{fig:CMD} upper left and lower left is the CMD constructed by all `pure' point sources and the panel (b) is constructed by the stars with $p_{\rm RGB}>0.9$. The panel (c) is a Hess diagram of the panel (a). The panel (d) shows the luminosity function of the stars with $p_{\rm RGB}>0.9$ (red; $i_0<23.5$) and all `pure' point sources (black; $i_0>23.5$). We overlay the BaSTI isochrone \citep{hidalgo2018,pietrinferni2021} with 10 Gyr and $[\mathrm{Fe/H}] = -1.90$ assuming the heliocentric distance of 958 kpc as the cyan solid lines (the metallicity and distance of Tri IV are estimated in the following section). The right panel of Figure \ref{fig:CMD} shows the residual Hess diagram and residual luminosity function (panel (b), (c), and (d) in the upper left figure minus those in the lower left figure). It is noted that the Tri IV field has stars with Tri IV and M33 halo stars, while the control M33 field has only M33 halo stars, so the difference shown as the color bar in the right figure of Figure \ref{fig:CMD} represents the true intensity of Tri IV.

From the panel (b) in the upper left figure and right figure of Figure \ref{fig:CMD}, we can confirm that Tri IV has a sparse RGB sequence. Besides, HB is confirmed around $((g-i)_0,i_0) \sim (0.25,25.5)$ in the panels (a) and (c). 
From the panels (a) and (c) in the lower right figure, the overdensity around ${\it i}_0 \sim 25.5$ can be seen. One of the main causes is the HB stars in the M33 smooth halo. In the control field subtracted luminosity function in the right figure of Figure \ref{fig:CMD}, we can see an overdensity of point sources around ${\it i}_0 \sim 25.5$ mag, so HB in the upper panel is more prominent than that in the lower panel. Finally, the faintest overdensity at ${\it i}_0 \sim 26.5$ is likely due to unresolved background galaxies that are not removed by the cuts described above.
Based on the diagnosis of the CMD, we can confirm the presence of a bulk of stellar populations in Tri IV.

\clearpage

\section{Photometric Properties of Triangulum IV}\label{section:properties}
\begin{table}[ht!]
 \caption{The photometric structural properties of Tri IV. The projected distances to M31 ($D_{\rm M31}$) and M33 ($D_{\rm M31}$) are calculated assuming the distances to M31 \citep[776 kpc][]{dalcanton2012a} and M33 \citep[859 kpc][]{degrijs2017}. The half-light radius $r_h$ expressed in arcmin units and {\it V}-band absolute magnitude $M_{\it V}$ are calculated assuming the distance of Tri IV (932 kpc).}
 \label{table:properties}
 \begin{tabular*}{\columnwidth}{@{}l@{\hspace*{100pt}}l@{}}
       \hline
       Property & Value \\
       \hline
       $\alpha$ & $01^{\rm h}28^{\rm m}38\fs93$ \\
       $\delta$ & $+30^{\circ}59'03\farcs6$ \\
       $D_{\odot}$ [kpc] & $932_{-43}^{+49}$\\
       $D_{\rm M31}$ [kpc] & $243_{-19}^{+30}$\\
       $D_{\rm M33}$ [kpc] & $75_{-40}^{+48}$\\
       $r_h$ [arcmin] & $6.5_{-1.3}^{+1.5}$ \\
       $r_h$ [pc] & $1749_{-425}^{+523}$ \\
       $M_{\it V}$ [mag] & $-6.39_{-0.52}^{+0.54}$ \\
       $\mu_{{\it V}}$ [mag arcmin$^{-2}$] & $29.72_{-0.10}^{+0.10}$\\
       $\epsilon$ & $0.21_{-0.14}^{+0.19}$ \\
       $\theta$ [degree] & $90_{-47}^{+56}$ \\
       $N_{\star}$ & $47_{-12}^{+13}$ \\
       ${\it E(B-V)}$ [mag]& $0.04$ \\
       $[\mathrm{Fe/H]}$ [dex] & $-1.90_{-0.08}^{+0.08}$\\
       \hline
 \end{tabular*}
\end{table}

The structural parameters of Tri IV are determined by the Markov Chain Monte Carlo (MCMC) method \citep{martin2009,martin2016a,martinez-delgado2021a}. This method estimates the properties of a dwarf galaxy by fitting a radial density profile, assuming that the observed data are member stars of the dwarf galaxy or contaminants. In this study, we assume the profile of a dwarf galaxy to be an exponential profile ($\rho_{\rm dwarf}$; see \citet{martinez-delgado2021a} for full equations), and the contamination population to be a uniform component ($\Sigma_{b}$). The parameters of the dwarf galaxy are its centroid ($\xi_0$, $\eta_0$), its exponential half-light radius ($r_h$), the position angle of its major-axis ($\theta$), its ellipticity ($\epsilon = 1 - b/a$, with $a$ and $b$ being the major- and minor-axis scale lengths, respectively), and the number of member stars ($N_\star$), which is tied to the contamination level.

From these assumptions, we construct a likelihood function,
\begin{equation}
\mathcal{L} = \prod_i^n \frac{\rho_{\rm dwarf} (r_i|\xi_0, \eta_0, r_h, \theta, \epsilon, N_\star) + \Sigma_{b}}{\int_\mathcal{A} \left[ \rho_{\rm dwarf} + \Sigma_{b} \right] d\mathcal{A}},
\end{equation}
where $\mathcal{A}$ is the region used for calculation and $r_i$ is an elliptical radius from the center of the dwarf galaxy such that:
\begin{equation}
\begin{split}
r_i = \left\{ \left[ \frac{1}{1-\epsilon}((\xi_i-\xi_0)\cos{\theta}-(\eta_i-\eta_0)\sin{\theta}) \right]^2 \right.\\
\left. + \left[ (\xi_i-\xi_0)\sin{\theta}+(\eta_i-\eta_0)\cos{\theta}\right]^2 \right\}^{1/2}.
\end{split}
\end{equation}
\nopagebreak[4]

We adopt broad and flat uniform priors (i.e., uninformative priors) for all parameters. We run the MCMC fitting, set to 100 walkers, 110,000 iterations, and first 100,000 burn-in using \texttt{emcee}. In Figure \ref{fig:CMD}, we confirm the existence of remaining background galaxies with $i_0 > 25.5$, so we conduct the MCMC fitting only for the RGB stars with $i_0 < 23.5$ and $p_{\rm RGB} > 0.9$, to ensure uniformity and to reduce the contamination as possible. We show the results of the MCMC fitting in Table \ref{table:properties}.

Based on the estimated value of the half-light radius, we also estimate the line-of-sight distance using the stars within 3 times the half-light radius of Tri IV. In Figure \ref{fig:CMD}, we confirm that the RGB sequence is sparse, so distance estimation is performed only for well-populated HB.

We perform the same method as \citet{cohen2018}. This method estimates the apparent magnitude of HB by fitting the model luminosity function to the observed one using MCMC. The model luminosity function consists of an exponential function that reproduces the RGB sequence, plus a double Gaussian function that reproduces RC and RGB bump. In this study, we do not change the form of the model function and likelihood themselves, but it should be noted that our double Gaussian model is to reproduce the RC population and the remaining background galaxies, because the background contaminants show a predominant peak and the RGB bump is buried (see, Figure \ref{fig:CMD}). We calculate the likelihood function using Equation (2) in \citet{cohen2018}, and apply the broad and flat uniform priors for all parameters. We run the MCMC method with 30 walkers and 110,000 iterations, where the first 100,000 iterations were thrown out to account for burn-in. 

The estimated apparent {\it g}-band magnitude of HB is $m_{{\rm HB},{\it g}}=25.40_{-0.03}^{+0.03}$. Using the formula for photometric transformation and the equation of absolute magnitude of HB in \citet{komiyama2018}, we estimate the distance of Tri IV. It is noted that the metallicity and HB color are required to estimate the absolute magnitude of HB. For the estimation of the metallicity of Tri IV, we perform the chi-square fitting of a BaSTI isochrone with age of 10 Gyr to the observed CMD, and we derive its metallicity with $[{\rm Fe/H }]\sim-1.90$. For the color of HB, we use the Gaussian-fitted estimation result ($(g-i)_0 \sim$ 0.25) to the observed color distribution. As a result, the absolute magnitude of HB is found to be $M_g \sim 0.55$ mag, and we find that the distance of Tri IV is found to be $932_{-43}^{+49}$ kpc. Assuming the heliocentric distance to M33 \citep[932 kpc][]{degrijs2017}, the 3-dimensional distance to Tri IV from the center of M33 is $75$ kpc, suggesting that Tri IV is within the virial radius of M33 \citep{patel2018}.

To characterize Tri IV, we estimate the absolute magnitude using the same method as \citet{martin2016a}. In this method, a probability density function (PDF) is constructed from the theoretical luminosity function, randomly sampled from that PDF, and the magnitude is determined by summing up the flux from the samples of the PDF. In this study, we construct the PDF using the BaSTI database \citep{hidalgo2018,pietrinferni2021} with age $= 10$ Gyr and $[{\rm Fe/H}]=-1.90$. From the constructed PDF, we sample the magnitude randomly. When sampling, if the sampled magnitude is $i_0<23.5$, we record it, and if the magnitude is fainter than that, we discard it. Once the number of recorded sampling reach up to $N_{\rm star}$, which is described in Table \ref{table:properties}, we add all sampled magnitudes to calculate the {\it g}- and {\it i}-band magnitude of Tri IV. We then convert these to the {\it V}-band using the conversion equation from \citet{komiyama2018}. To measure the average luminosity and the standard deviation, we repeat this procedure 10,000 times, and obtain $M_{\it V} = -6.39_{-0.52}^{+0.54}$ for a distance of $D_{\odot} = 932_{-43}^{+49}$ kpc where both distance and luminosity uncertainties are included. The estimated absolute magnitude is shown in Table \ref{table:properties}.

\section{Discussion and Conclusions}\label{section:conclusions}

\begin{figure*}[ht!]
\includegraphics[width=2\columnwidth]
{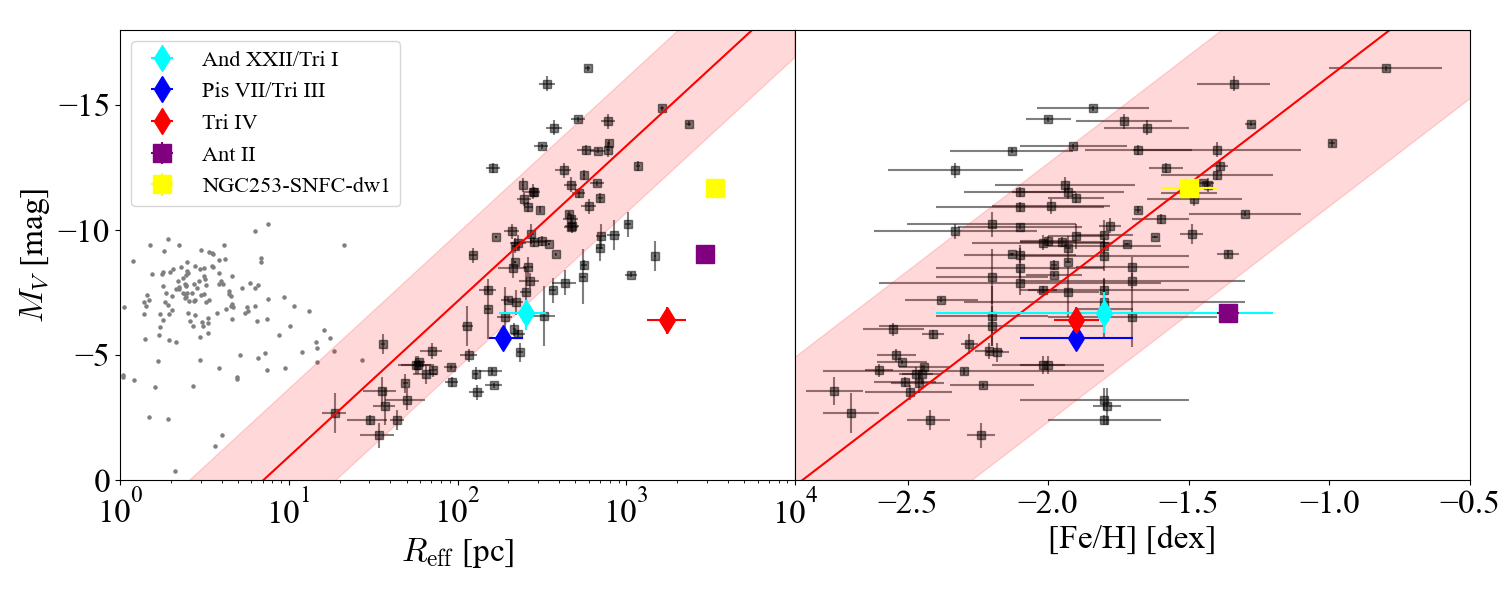}
\caption{Left: Absolute {\it V}-band magnitude ($M_V$) v.s. effective radius ($R_{\rm eff}$) plane. Black squares show the dwarf galaxies and gray dots show the globular clusters \citep{harris1996,mcconnachie2012}. The red diamond is Tri IV, and the blue and cyan ones are Andromeda XXII/Triangulum I (And XXII/Tri I) and Pisces VII/Triangulum III (Pis VII/Tri III), which are the satellites of M33. The purple and yellow squares represent Antlia II (Ant II) and NGC253-SNFC-dw1. The red line represents the distribution of black squares, and the shaded region represents the 1-sigma range. Right: Absolute {\it V}-band magnitude ($M_V$) v.s. metallicity. Black squares show the dwarf galaxies and diamonds show the satellites of M33. The red diamond illustrates Tri IV. The red line represents the distribution of black squares, and the shaded region represents the 1-sigma range.\label{fig:properies}}
\end{figure*}

We have carried out deep imaging of the western region of M33 using Subaru/HSC. This observation reaches down to $i_0\sim 27$ mag. Based on the {\it NB515} information, we have extracted M33 RGB stars effectively, and succeeded in detecting an overdensity, Triangulum IV (Tri IV). The CMD of Tri IV exhibits sparse RGB stars and the AGB bump. HB stars have been contaminated by the unresolved background galaxies, despite the removal of the background galaxy in the color-color space. However, the luminosity function of Tri IV clearly shows the peak at $i_0 \sim 25.5$ which corresponds to the HB population of Tri IV.

We have characterized the photometric properties of Tri IV, and we find that Tri IV is located $\sim 75$ kpc ($\sim 243$ kpc) from the center of M33 (M31), so Tri IV is within the virial radius of M33 \citep[$\sim 160$ kpc;][]{patel2018}. In addition, our estimated values suggest that Tri IV is dominated by a metal-poor population, and it has a large radius and low surface brightness. Its faint and diffuse structure is a major reason why Tri IV has not been discovered so far. 

The diffuse nature of Tri IV suggests that Tri IV may be an ultra-diffuse galaxy or a dynamically heated dwarf galaxy. The left panel of Figure \ref{fig:properies} shows the luminosity-size diagram of the dwarf galaxies \citep{mcconnachie2012} and globular clusters \citep{harris1996}. The diamonds show the M33 satellites with Tri IV shown in red diamonds. The red line represents the distribution of black squares, and the shaded region represents the 1-sigma range. In this panel, Tri IV is near the dwarf galaxy sequence, but its characteristic size makes it more than 1-sigma deviated from typical dwarf galaxies. Another example of such large-sized galaxies is Antlia II \citep[AntII;][]{torrealba2019} shown as the purple square in Figure \ref{fig:properies}, and the similarity between Ant II and Tri IV is that both are characterized by low surface brightness. \citet{reed2023} suggested that Ant II have been disrupted by tidal forces of the Galaxy. In addition, one of the dwarf galaxies, NGC~253-SNFC-dw1 which is the ultra-diffuse galaxy in NGC~253 shown as the yellow square in Figure \ref{fig:properies}, has been suggested to be in the process of passing through the apocenter of its orbit due to its asymmetric structure \citep{okamoto2024}. In Figure \ref{fig:SpatialDistribution}, a clump of RGB stars can be seen on the north side of Tri IV, and this structure may be stripped from the main body of Tri IV. Therefore, Tri IV may have been similarly stretched by tidal interactions with M33. To confirm this disruption scenario, future follow-up spectroscopic observations for stars in Tri IV are required using Subaru/Prime-Focus Spectrograph \citep{takada2014}

The right panel of Figure \ref{fig:properies} illustrates the luminosity-metallicity diagram of the dwarf galaxies \citep{mcconnachie2012}. The diamonds represent the M33 satellites, and Tri IV shows red diamonds. The red line represents the distribution of black squares, and the shaded region represents the 1-sigma range. In this panel, Tri IV is found to have properties consistent with those of the typical dwarf galaxies in the 1-sigma range. Therefore, although the size of Tri IV is found to be exceptionally large, its photometric metallicity is almost the same as that of the typical dwarf. Furthermore, the location of Tri IV within the normal dwarf systems in the $M_{\it V}$ vs $[{\rm Fe/H}]$ suggests that the stellar component of Tri IV has not lost much stellar mass, although it has been transformed into an extended and very low surface brightness distribution, presumably due to some dynamical heating. This is consistent with models such as \citet{reed2023}.

\citet{patel2018} predicted that there are about 6 satellites brighter than $M_V<-6$ within the virial radius of M33. So far, only one satellite galaxy (And XXII/Tri I) and one satellite candidate (Pis VII/Tri III) have been detected in M33. Even if Tri IV is confirmed to be a dwarf satellite of M33, the number is still insufficient. However, the observational region to detect the satellite galaxies in M33 is smaller than that predicted in \citet{patel2018}, and if limited to the Pan-Andromeda Archaeological Survey \citep{ibata2014,mcconnachie2018} region, the number of satellite galaxies in M33, including Tri IV, is almost consistent with the prediction in \citet{patel2018}. So far, there are few satellite galaxies in M33, which have been attributed to the tidal interaction with M31. However, the number of satellite galaxies detected by observations is equal to the theoretically predicted number, so M33 may not be affected by the tidal interaction of M31 and may be in a first infall, which is suggested using a combination of the proper motion of M33 and inference from cosmological simulations \citep[e.g.,][]{patel2017a,vandermarel2019}.

We acknowledge support in part from MEXT Grant-in-Aid for Scientific Research (No.~JP18H05437 and JP21H05448 for M.C., No.~JP21K13909 and JP23H04009 for K.H., and No.~JP22K14076 for T.K.). This work was partially supported by Overseas Travel Fund for Students (2023) of Astronomical Science Program, the Graduate University for Advanced Studies, SOKENDAI. E.N.K.\ acknowledges support from NSF CAREER grant AST-2233781. CF and RFGW are grateful for support through the generosity of Eric and Wendy Schmidt, by recommendation of the Schmidt Futures program.
Data analysis was in part carried out on the Multi-wavelength Data Analysis System operated by the Astronomy Data Center (NAOJ/ADC) and the Large-scale data analysis system co-operated by the ADC and Subaru Telescope, NAOJ. Based in part on data collected at Subaru Telescope and obtained from the SMOKA, which is operated by the NAOJ/ADC. 

\vspace{5mm}
\software{emcee \citep{foreman-mackey2013},
          astropy \citep{theastropycollaboration2013},
          Matplotlib \citep{hunter2007},
          numpy \citep{vanderwalt2011},
          corner \citep{foreman-mackey2016}}


\bibliography{TriIV}{}
\bibliographystyle{aasjournal}

\end{document}